\documentclass{ws-ijtaf}
\usepackage{graphicx} % Include figure files
\usepackage{bm} % bold math
\begin{document}

\markboth{Luca Capriotti}
{\large Least Squares Importance Sampling for
        Libor Market Models}% Force line breaks with \\

\title{\large Least Squares Importance Sampling for
        Libor Market Models}% Force line breaks with \\
\vspace{-0.5cm}

\author{\normalsize Luca Capriotti}
\address{
Global Modelling and Analytics Group, Investment Banking Division, Credit Suisse Group\\
One Cabot Square, London, E14 4QJ, United Kingdom \\
\email{luca.capriotti@credit-suisse.com} }

\maketitle

\begin{abstract}
A recently introduced Importance Sampling strategy based on a least squares
optimization is applied to the Monte Carlo simulation of Libor Market Models.  
Such Least Squares Importance Sampling (LSIS) allows the automatic optimization
of the sampling distribution within a trial class by means of a quick 
presimulation algorithm of straightforward implementation. 
With several numerical examples we show that LSIS can be extremely effective in 
reducing the variance of Monte Carlo estimators often resulting, especially when combined 
with stratified sampling, in computational speed-ups of orders of magnitude. 
\vspace{1pc}
\end{abstract}

\section{Introduction}

The level of sophistication of the models employed by investment firms
for pricing derivative securities is dramatically increasing 
in the continuous search for a possible edge against competitors.  As a result, 
most of the models used in practice is
too complex to be treated by analytic or deterministic numerical
methods, and Monte Carlo
simulation becomes more often than ever the only 
feasible means of pricing and hedging.

The main limitation of Monte Carlo simulations is their computational cost.
In fact, being stochastic in nature,
their outcome is always affected by a statistical error, that can
be generally reduced  to the desired level of accuracy by
iterating the calculation for long enough time.  This comes with a high computational cost
as such statistical uncertainties, all things being equal, are
inversely proportional to the {\rm square root} of the number of
statistically independent samples. Hence, in order to reduce the
error by a factor of 10 one has to spend 100 times as much
computer time. For this reason, to be used on a trading floor,
Monte Carlo simulations often require to be run on large parallel
computers with a high financial cost in terms of hardware,
infrastructure, and software development.

Several approaches to speed-up Monte Carlo calculations, such as Antithetic Variables,
Control Variates, and Importance Sampling, have been proposed over
the last few years \cite{GlassMCbook}. These techniques aim at 
reducing the variance per Monte Carlo observation so that a given
level of accuracy can be obtained with a smaller number of
iterations. In general, this can be done by exploiting some
information known {\em a priori} on the structure of the problem at
hand, like a symmetry property of the Brownian paths (Antithetic
Variables), the value of a closely related security (Control
Variates), or the form of the statistical distribution of the
random samples (Importance Sampling). Antithetic Variables and
Control Variates are the most commonly used variance reduction
techniques, mainly because of the simplicity of their
implementation, and the fact that they can be accommodated  in an
existing Monte Carlo calculator with a small effort. However,
their effectiveness varies largely across applications, and is
sometimes rather limited \cite{GlassMCbook}.

On the other hand, Importance Sampling techniques, although
potentially more powerful, have not been employed much in
professional contexts until recently.  This is mainly
because they generally involve a bigger implementation
effort. Moreover, when used improperly,
Importance Sampling can increase the variance of the Monte Carlo estimators, 
thus making its
integration in an automated environment more delicate.
Nonetheless, the potential efficiency gains at stake are so large
that the interest in finding efficient Importance Sampling schemes
is still very high.

The idea behind Importance Sampling is to reduce the statistical
uncertainty of a Monte Carlo calculation by focusing on the most
important sectors of the space from which the random samples are
drawn.   Such regions critically depend on both the random
process simulated, and  the structure of the security priced.
For instance, for a deep out-of-the money Call option \cite{Hull},
the payoff sampled is zero for most of the iterations of a Monte
Carlo simulation. Hence, simulating more samples with positive
payoff reduces the variance. This can be done by changing the
probability density from which the samples are drawn, and
reweighing the payout function by the appropriate likelihood-ratio
(Radon-Nikodym derivative) in order to produce an unbiased result
of the original problem \cite{GlassMCbook}.

Most of the work in Importance Sampling methods for security pricing has been done in a
Gaussian setting
\cite{reider,BoyleBroadieGlass97,VasDuf98,GlassImportSampl99,
GlassImportSampl99hjm,SUFU00,SUFU02,Arouna03,GuasRob06} such the one arising
from the simulation of a diffusion process. In this framework,
Importance Sampling is achieved by modifying the drift term of the
simulated process in order to drive the Brownian paths towards the
regions that are the most important for the evaluation of the
security. For instance, for the  Call option above, this can be
obtained by increasing the drift term up to a certain
optimal level \cite{reider,BoyleBroadieGlass97}. 
The various approaches proposed in the literature, essentially differ in the
way in which such change of drift is found, and can be roughly
divided into two families depending on the strategy adopted. The
first strategy, common to the so-called adaptive Monte Carlo
methods \cite{VasDuf98,SUFU00,SUFU02,Arouna03}, aims to determine
the optimal drift  through stochastic optimization techniques that
typically involve an iterative algorithm. On the other hand, the
second strategy, proposed in a remarkable paper by Glasserman,
Heidelberger, and Shahabuddin (GHS) \cite{GlassImportSampl99},
relies on a deterministic optimization procedure that can be
applied for a specific class of payouts. 
%This approach, although
%approximate, turns out to be very effective for several pricing
%problems, including the simulation of a single factor
%Heath-Jarrow-Morton model \cite{GlassImportSampl99hjm}, and
%portfolio credit scenarios \cite{GlassLi05}.

In a recent paper \cite{LC_lsis06}, we introduced the Least Squares Importance Sampling
(LSIS) technique, as an alternative and flexible variance reduction strategy for
Monte Carlo security pricing. This approach, originally proposed in Physics
for the optimization of quantum mechanical wave functions of
correlated electrons \cite{umrigar98}, was shown in Ref.~\cite{LC_lsis06} to provide 
an effective tool also for financial applications.
In LSIS the determination of the optimal drift
-- or more in general of the most important regions of the sample
space -- is formulated in terms of a least squares minimization. 
This technique can be easily implemented and included
in an existing Monte Carlo code, and simply relies on a
standard least square algorithm for which several optimized
libraries are available. 

In this paper we apply the LSIS strategy to the simulation of a multi-factor 
Libor Market Model, and test its effectiveness on a variety of contracts. 
In addition, to further increase the computational efficiency we combine LSIS
with stratified sampling \cite{HammersleyHandscomb64}. The resulting variance reduction
strategy is shown to be quite effective in a variety of cases, providing computational speed-ups 
of up to two orders of magnitude.  

In the following Section, we begin by 
discussing the simulation setting to which we apply the LSIS strategy.
Then in Section \ref{importance}  we review the main ideas behind 
Importance Sampling, and the principal approaches proposed in the financial literature. 
The rationale of LSIS is discussed in Section \ref{leastsquares} 
together with the essential implementation details, and in Section
\ref{ss} we illustrate how to combine LSIS with stratified sampling.
Sections \ref{lmm} and \ref{numres} discuss the Libor Market Model setting,
and present the numerical results obtained with
LSIS in this case.  Finally, we draw our conclusions in Section \ref{conclusions}.

\begin{figure}
\vspace{-14mm}
\center
\includegraphics[width=0.83\textwidth]{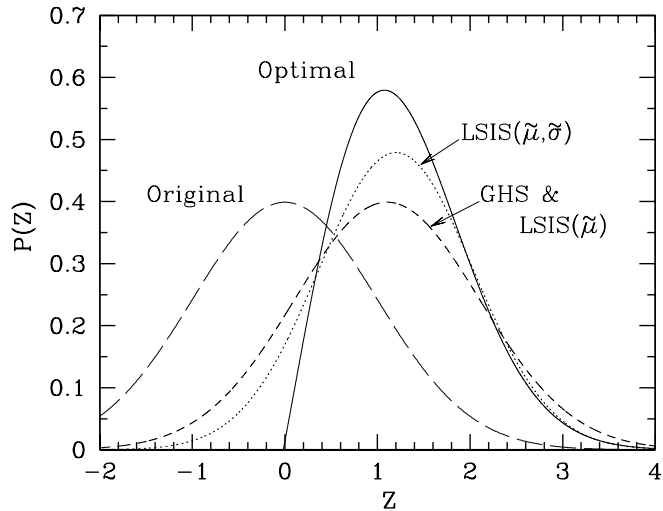}
\vspace{-10mm} \caption{\label{figcall} Sampling probability
density functions for a European Call option (\ref{call}) with $T = 1$, $r =
0.05$, $\sigma = 0.3$, $X_0 = K = 50$ as obtained with LSIS
[optimizing just the drift, LSIS$(\tilde\mu)$, and both the drift and
the volatility, LSIS$(\tilde\mu,\tilde\sigma)$], and the saddle point approximation
of Ref.~\protect\cite{GlassImportSampl99} (GHS). On this scale the results for LSIS$(\tilde\mu)$ and GHS are
indistinguishable. The original (\ref{gaussmulti}) and the optimal
(\ref{zerovar}) sampling densities are also shown for comparison.
}
\end{figure}

\section{The Setting}

Although the variance reduction technique we discuss in this paper can be applied 
to a variety of financial problems, in 
the following we will focus on pricing applications that involve the simulation 
of  multi-dimensional diffusions of the form
\begin{equation}\label{sde1}
dX(t) = \mu(X(t),t)\,dt+\sigma(X(t),t)\,dW_t~.
\end{equation}
Here the process $X(t)$ and the drift $\mu(X,t)$ are both
$L$-dimensional real vectors, $W_t$ is a $N$-dimensional standard
Brownian motion, and the volatility, $\sigma(X,t)$, is a $L\times N$
real matrix. We will consider the problem of estimating
the value at time $t=0$, of contracts depending 
on the path followed by $X(t)$ within a certain interval
$[0,T]$. This is given by the expectation value under the risk neutral
probability measure, $P$ \cite{HarrKreps} of the (discounted) payout
functional $G[X(T)]$
\begin{equation}
V = E_P\left[ G[X(T)] \right]~.
\end{equation}

Continuous time processes of the form (\ref{sde1}) are
typically simulated by sampling $X(t)$ on a discrete grid of
points, $0=t_0 < t_1 < \ldots < t_M =T$, by means, for instance,
of a Euler scheme \footnote{The use of other discretization
schemes does not alter the present discussion.}
\begin{equation}\label{sdeeuler}
X_{i+1} = X_i + \mu(X_i,t)\,\Delta t_i+\sigma(X_i,t)\,\sqrt{\Delta
t_i}\, \tilde Z_{i+1}~,
\end{equation}
where $X_i = X(t_i)$, $\Delta t_i = t_{i+1}-t_i$, and $\tilde Z_{i+1}$ is
a $N$-dimensional vector of independent standard normal variates.
In this representation, each discretized path for the vector
process $X(t)$ can be  put into a one to one correspondence with a
set of $d = N\times M$ independent standard normal variables $Z$.
As a result, the original problem of evaluating the expectation value of
a functional of the realized path of the process $X(t)$ can be
formulated as the calculation of expectation values of the form
\begin{equation}\label{expectproblem}
V = {E}_P\left[ G(Z) \right] = \int \hspace{-1mm} dZ \,\, G(Z)\,
P(Z)~,
\end{equation}
where  $G(Z) = G(Z_1,\ldots,Z_d)$ is the scalar function
obtained by discretizing the payout functional $G[X(T)]$ 
on a mesh of $d$ sampling points, 
and the density is given by a $d$-dimensional standard normal distribution
\begin{equation}\label{gaussmulti}
P(Z) = N(0, I_d) \equiv {(2\pi)^{-d/2}} \,\, e^{ -Z^2/2}~,
\end{equation}
where $Z^2 = Z\cdot Z$.
For instance, for
the familiar Call option in the Black-Scholes framework
\cite{Hull} one has $d=1$, $P(Z) = (2\pi)^{-1/2}
\exp{(-Z^2/2)}$ and
\begin{equation}\label{call}
G(Z) =
e^{-rT}\left(X_0\exp{\left[\left(r-\frac{\sigma^2}{2}\right) T +
\sigma \sqrt{T} Z\right]}-K\right)^+
\end{equation}
 where $r$ is the risk-free interest rate, $\sigma$ is the
volatility, $X_0$ and $K$ are respectively the spot and strike
price, and $T$ the maturity of the option.

Whenever the dimension $d$ of the state variable $Z$ is large (say
$d \gtrsim 5$) standard numerical quadrature approaches become
highly inefficient, and Monte Carlo methods are the only feasible
route for estimating expectation values of the form
(\ref{expectproblem}). To do so, one interprets
Eq.~(\ref{expectproblem}) as a weighted average of the payout
function $G(Z)$ over the possible configurations $Z$ with weights
given by the probability density $P(Z)$. This immediately
leads to the simplest (and crudest) Monte Carlo estimator which is
obtained by averaging the payout function over a sample of $N_p$
{\em independent} values of the random variable $Z$ generated
according to the probability density $P(Z)$,
\begin{equation}\label{crude}
V \simeq \bar V = \frac{1}{N_p} \sum_{i=1}^{N_p} G(Z_i) ~~~~~~ Z_i
\sim P(Z)~.
\end{equation}
In particular, the central limit theorem \cite{CLT} ensures that,
for big enough samples, the values of the estimator $\bar V$ are
normally distributed around the true value, and converge for $N_p
\to \infty$ towards $V$ namely
\begin{equation}\label{sqrt}
V \simeq \frac{1}{N_p} \sum_{i=1}^{N_p} G(Z_i) \pm
\frac{\Sigma}{\sqrt{N_p}}~,
\end{equation}
where $\Sigma^2 =
E_P\left[G(x)^2\right]-E_P\left[G(x)\right]^2$ is
the variance of the estimator and can be similarly
approximated by
\begin{equation}
\Sigma^2 \simeq  \frac{1}{N_p} \sum_{i=1}^{N_p} \left(G(Z_i)
- \bar V \right)^2~.
\end{equation}
Although Eq.~(\ref{sqrt}) ensures the convergence of the Monte
Carlo estimator to the expectation value (\ref{expectproblem}),
its practical utility depends on the magnitude of the variance,
$\Sigma^2$.
Indeed, the square root convergence in (\ref{sqrt}),
implies that the number of replications $N_p$ that are
(asymptotically) necessary to achieve a given level of accuracy is
proportional to the variance of the estimator
\footnote{In particular, the Monte Carlo integration
becomes unfeasible if the variance of the estimator diverges, 
giving rise to the so-called {\em
sign-problem} instability. Although this problem is the crux of
Monte Carlo simulations in several branches of the Physical
Sciences, see, e.g., S. Sorella and L. Capriotti, Physical Review
B {\bf 61}, 2599 (2000), this issue does not usually affect
financial contexts.}.
Roughly speaking,
such quantity is relatively small whenever the function $G(Z)$ is
approximately constant over the region of values of $Z$ that is
represented the most among the random samples, i.e., the region
that contains most of the probability mass of $P(Z)$. This is
generally not the case for most of the pricing problems
encountered in practice, and the calculation of accurate estimates
of the expectation value (\ref{expectproblem}) may require large
sample sizes $N_p$, thus becoming computationally demanding.

\section{Importance Sampling}
\label{importance}

The key observation underlying Importance Sampling is that
the choice of extracting the random variable $Z$
according to the probability density $P(Z)$ in order to sample
stochastically Eq.~(\ref{expectproblem}), although
natural, is by no means the only possible one. Indeed, the Monte
Carlo integration can be performed by sampling an arbitrary
probability density $\tilde P(Z)$ provided that the integral
is suitably reweighed. In fact, using the identity
\begin{equation}
\int \hspace{-1mm} dZ \,\, G(Z)\, P(Z) = \int \hspace{-1mm} dZ
\,\, \frac {G(Z) P(Z)}{\tilde P(Z)}\, \tilde P(Z)~,
\end{equation}
an alternative estimator of the expectation value
(\ref{expectproblem}) is readily found as
\begin{equation}\label{isamp}
V \simeq \tilde V = \frac{1}{N_p} \sum_{i=1}^{N_p} W(Z_i) \,
G(Z_i) ~~~~~~ Z_i \sim \tilde P(Z)~,
\end{equation}
with the weight function given by $W(Z) = P(Z)/\tilde P(Z)$.
The variance of the new Monte Carlo estimator reads
\begin{equation}\label{varnew}
\tilde \Sigma^2 = \int \hspace{-1mm} dZ \,\, \left(W(Z)\,G(Z) -
V\right)^2\,\tilde P(Z)
\end{equation}
and critically depends on the choice of the sampling probability
density $\tilde P(Z)$. For non-negative functions $G(Z)$, the
optimal choice of $\tilde P(Z)$ is the one for which $\tilde
\Sigma$ vanishes, namely:
\begin{equation}\label{zerovar}
 P_{\rm opt}(Z) = \frac{1}{V} \, G(Z) P(Z)~.
\end{equation}
In fact, the Monte Carlo estimator corresponding to such {\em
optimal sampling density} reads
\begin{equation}
\tilde V \simeq \frac{1}{N_p} \sum_{i=1}^{N_p} W(Z_i) G(Z_i) =
\frac{1}{N_p} \sum_{i=1}^{N_p} V  ~,
\end{equation}
leading to a constant value $V$ on each Monte Carlo replication,
and resulting therefore in zero variance \footnote{It is possible to show
\cite{numrec} that, when $G(Z)$ does not have a definite sign, the
optimal sampling density has the similar form $P_{\rm opt} = |G(Z)|
P(Z)/V$, although in this case the resulting variance is not zero.}.
Unfortunately, such a choice is not really viable as the
normalization constant, $V$, is the expectation value
(\ref{expectproblem}) we want to calculate in the first place.
Nevertheless, this observation provides the useful indication that the sampling density $\tilde{P}(Z)$,
modulus a normalization, should be as close as possible to the product of the payout $G(Z)$ and the
original multi-variate Gaussian distribution (\ref{gaussmulti}).

In this respect, Importance Sampling strategies generally choose
a family of trial probability densities, $\tilde P_{\theta}(Z)$ -- 
depending on a set of $N_\theta$ real parameters $\theta = (\theta_1, \theta_2,
\ldots, \theta_{N_\theta})$~ -- and aim at determining the one that
minimize the variance of the estimator (\ref{varnew}) within the class.
In particular,  Importance Sampling methods
in security pricing generally try
to guide the sampled paths towards the most important regions of the configuration space (i.e.,
where the contribution of the integrand is the largest), by means
of a change of the drift terms of the process (\ref{sde1}) or
(\ref{sdeeuler}). The corresponding  trial probability density
reads
\begin{equation}\label{trialgauss}
\tilde P_{\tilde\mu}(Z) = (2\pi)^{-d/2}  \,\, e^{-(Z-\tilde\mu)^2/2}~,
\end{equation}
 where $\tilde \mu$ is a
$d$-dimensional vector, and the weight function, as also expected
from the Girsanov theorem \cite{Musiela}, is
\begin{equation}
W_{\tilde\mu} (Z) = \exp\left[- \tilde\mu\cdot Z + \tilde\mu^2/2 \right]~.
\end{equation}

A variety of approaches for the determination of the drift vector
$\tilde\mu$ minimizing the variance of the estimator
(\ref{varnew}) has been recently proposed in the literature
\cite{VasDuf98,GlassImportSampl99,
GlassImportSampl99hjm,SUFU00,SUFU02,Arouna03}. These can be
roughly classified into two families depending on the strategy
adopted. 

The first strategy, common to the so-called adaptive
Monte Carlo methods, is based on a stochastic minimization of the variance. Such
minimization differs in details in the various methods but always
involves an iterative procedure,  to be performed in a preliminary
Monte Carlo simulation.

In particular, Su and Fu \cite{SUFU00,SUFU02}, building upon
previous work by Vazquez-Abad and Dufresne \cite{VasDuf98}, used a
gradient-based stochastic approximation, dubbed infinitesimal
perturbation analysis, in order to estimate the optimal {\em
uniform shift} of the drift for the diffusion (\ref{sdeeuler}),
minimizing the variance of the estimator (\ref{varnew}). In the
notation of this Section, this translates in working with a trial
density of the form (\ref{trialgauss}) where the drift vector
$\tilde\mu$ has components all equal to a single optimization
parameter. The improvement of this method with respect to the one
of Ref.~\cite{VasDuf98}, is that the minimization is carried out
under the original probability measure,
while in the latter the minimization was formulated under
the trial probability measure. As a result, the stochastic
minimization applies also for non differentiable payout, thus
making the approach more general. The application of this
technique to partial average Asian options in a Black-Scholes
market, and to Caplets under the Cox-Ingersoll-Ross model provides
significative variance reductions \cite{SUFU00,SUFU02}.

Along similar ideas, Arouna \cite{Arouna03} has recently proposed
a different stochastic optimization method for the determination
of  the optimal sampling density (\ref{trialgauss}). Here, in
contrast to the previous approach, all the components of the drift
vector are independently optimized. The method relies on  a
truncated version of the Robbins-Monro algorithm that is shown to
converge asymptotically to the optimal drift, and to provide an
effective variance reduction in a variety of cases. 

On the other hand, the alternative strategy for the optimization of the trial density
(\ref{trialgauss}), proposed by Glasserman, Heidelberger, and
Shahabuddin \cite{GlassImportSampl99}, relies on a saddle point
approximation to minimize the variance of the estimator
(\ref{varnew}), or equivalently of its second moment (in the
original measure)
\begin{equation}\label{secondmom}
m_2(\tilde \mu) = \int \hspace{-1mm} dZ \,\, W_{\tilde \mu}(Z)\,G(Z)^2 \,P(Z)~.
\end{equation}
In fact, if the payout function $G(Z)$ is positive definite, by
defining $F(Z) = \log G(Z)$ one can approximate
Eq.~(\ref{secondmom}) with the zero-order saddle point expansion
\begin{eqnarray}
&&(2\pi)^{-d/2} \int \hspace{-1mm}dZ \,\exp\left[2F(Z) - \tilde\mu \cdot Z + \tilde\mu^2 /2  - Z^2/2\right] \nonumber \\
&\simeq& C\,\exp \Big[\max_{Z}\left(2F(Z) - \tilde\mu \cdot Z +  \tilde\mu^2 /2  -
Z^2/2 \right)\Big]~, \nonumber
\end{eqnarray}
where $C$ is a constant. As a result, within this approximation,
the problem of determining the optimal change of drift boils down
to finding the vector $\mu$ such that
\begin{equation}
\max_{Z}\left(2F(Z) - \tilde\mu \cdot Z +  \tilde\mu^2 /2  -
Z^2/2 \right)
\end{equation}
is minimum.  It is easy to show that this is obtained by choosing
$\tilde \mu^\star = Z^\star$ where $Z^\star$ is the point that
solves the optimization problem
\begin{equation} \label{optimization}
\max \left(F(Z) - Z^2/2\right)~,
\end{equation}
or equivalently, for which the payout times the original
density, $ G(Z) P(Z)$, is maximum, i.e., $Z^\star$
corresponds to the maximum of the optimal sampling density,
Eq.~(\ref{zerovar}).  The simplest interpretation of the saddle
point approach is therefore that it approximates the zero variance
density by means of a normal density with the same mode and
variance.

This approach has been recently generalized to the continuous time
in the Black-Scholes framework in a recent work by Guasoni and Robertson \cite{GuasRob06}.
This formulation allows one to express the problem of the determination
of the optimal drift in terms of a one-dimensional variational
problem, and the solution of a Euler Lagrange equation.

The saddle point approach can be expected to be particularly
effective in reducing the variance of the Monte Carlo estimator
whenever the log payout function $F(Z)$ is close to be linear in
the portion of the configuration space where  most of the
probability mass of $P(Z)$ lays. 
However, whenever
the optimal sampling probability (\ref{zerovar}) cannot be
accurately represented by a single Gaussian with the same mode and
variance, the saddle point approximation is less beneficial. In
particular,  this approach turns out to be less
effective whenever the structure of the payout function $G(Z)$ is
such that the optimal sampling density (\ref{zerovar}) has a
width  which is very different from the one of the original
density, or is multi-modal.

In the following Section we describe an alternative least squares
strategy that is straightforward to implement and flexible enough
to be applied in a generic Monte Carlo setting. Indeed, the Least
Squares Importance Sampling (LSIS) is not limited to the
determination of the optimal change of drift in a Gaussian
model. Instead, it can be applied to any Monte Carlo simulation
provided that a reasonable guess of the optimal sampling density
is available. For this reason, in the next Section we will
momentarily leave the Gaussian framework, and we will describe the
rationale of LSIS in a more general setting.

\begin{table*}
\caption{\label{tablecall} Variance reductions (\ref{varred}) obtained
with different Importance Sampling strategies. Comparison between LSIS, the adaptive
Robbins-Monro (RM) algorithm (as quoted in Ref.~\protect\cite{Arouna03}),
and the saddle point approach of Ref.~\protect\cite{GlassImportSampl99}
(GSH): price of a European Call option on a lognormal asset
(\ref{call}) for different values of the volatility $\sigma$,
and of the strike price $K$. The parameters used are $r = 0.05$,
$X_0=50$, $T=1.0$, and the number of simulated paths is 1,000,000
for Crude MC, LSIS and GHS, 50,000 for RM. Results for LSIS
obtained by optimizing  the drift only [LSIS$(\tilde\mu)$], and
both the drift and the volatility [LSIS$(\tilde\mu,\tilde\sigma)$]
are reported. The uncertainties are reported in
parentheses.}
\center
\begin{tabular}{cccccc}
$\sigma$ & $K$ & LSIS($\tilde\mu$)   & LSIS$(\tilde\mu,\tilde\sigma)$ & RM      & GHS     \\ \hline
0.1      &  30 &  104(1)             & 1700(100)                      & 112(4)  & 100(1)  \\
         &  50 &  7.8(1)             & 15(1)                          & 7.8(4)  & 7.8(1)  \\
         &  60 &  33.5(5)            & 84(5)                          & 31(2)   & 33.5(5) \\
0.3      &  30 &  16.4(1)            & 51(1)                          & 16.8(4) & 14.8(2) \\
         &  50 &  9.9(5)             & 27(1)                          & 11(2)   & 9.9(1)  \\
         &  60 &  15.6(1)            & 35(1)                          & 15.2(4) & 14.2(1) \\
\end{tabular}
\end{table*}

\begin{table*}
\caption{Same as Table \ref{tablecall} for a
European Put option.}
\label{tableput}
\center
\begin{tabular}{cccccc}
$\sigma$ & $K$ & LSIS($\tilde\mu$)   & LSIS$(\tilde\mu,\tilde\sigma)$ & RM      & GHS     \\ \hline
0.1      & 40  & 435(6)            & 571(9)                         & 350(24)    & 435(6)  \\
         & 50  & 8.8(1)            & 25(2)                          & 9.6(4)     & 9.1(1) \\
         & 60  & 5.9(1)            & 17(1)                          & 6.3(4)     & 5.9(1) \\
0.3      & 30  & 41(1)             & 69(2)                          & 38(4)      & 40.8(5) \\
         & 50  & 5.8(1)            & 16.5(5)                        & 6.2(4)     & 5.8(1) \\
         & 60  & 4.9(1)            & 13.9(2)                        & 4.8(4)     & 4.4(1) \\
\end{tabular}
\end{table*}

\section{Least Squares Importance Sampling}
\label{leastsquares}

A practical approach to the search of an effective Importance
Sampling density can be formulated in terms of a non-linear
optimization problem. To this purpose, let us consider the family of
trial probability densities, $\tilde P_{\theta}(Z)$. 
The variance of the estimator
corresponding to $\tilde P_{\theta}(Z)$, Eq.~(\ref{varnew}), can
be written in terms of the {\em original} probability density
$P(Z)$ as
\begin{equation}\label{varest}
\tilde \Sigma_\theta^2 = E_P \left[W_\theta(Z) G^2(Z) \right] -
E_P \left[G(Z) \right]^2~,
\end{equation}
with $W_\theta(Z) = P(Z)/\tilde P_\theta(Z)$. Hence, the optimal
Importance Sampling density within the family $\tilde P_\theta(Z)$
is the one for which the
latter quantity, or equivalently the second moment
(\ref{secondmom}) or
\begin{equation}\label{second}
E_P \left[W_\theta(Z) G^2(Z) \right]~,
\end{equation}
is minimum. The crucial observation is that the Monte Carlo
estimator of this quantity,
\begin{equation}\label{minimizer}
m_2(\theta) \simeq \frac{1}{N_p^\prime} \sum_{i=1}^{N_p^\prime} \left( W_\theta(Z_i)^{1/2}
G(Z_i)\right)^2 \,\,\,\,\, Z_i\sim P(Z)~,
\end{equation}
can be interpreted as a non-linear least squares fit of a set of
$N_p^\prime$ data points $(x_i,y_i)$ with a function $y = f_\theta(x)$
parameterized by $\theta$, with the correspondence $y_i \to 0$,
$x_i \to Z_i$, and $f_\theta(x) \to W_\theta(Z)^{1/2} G(Z)$.
The latter is a standard problem of statistical analysis that can
be tackled with a variety of robust and easily accessible
numerical algorithms, as the so-called Levenberg-Marquardt method
\cite{numrec}.

Alternatively, to improve the numerical stability of the
least-squares procedure, it is convenient in some situations
to minimize, instead of (\ref{second}),  the pseudo-variance
\begin{eqnarray}\label{pseudovariance}
S_2(\theta) &=& E_P\left[ \left( W_\theta(Z)^{1/2}
G(Z) - V_T \right)^2\right] \nonumber \\
&&\simeq\frac{1}{N_p^\prime}\sum_{i=1}^{N_p^\prime} \left( W_\theta(Z_i)^{1/2}
G(Z_i) - V_T \right)^2 \,
\end{eqnarray}
where the constant $V_T$ is a guess of the option value. Indeed, the minimization
of (\ref{pseudovariance}) is equivalent to the one of the real variance of the estimator
(\ref{varest}) as
\begin{equation}
S_2(\theta) = \tilde \Sigma_\theta^2 + \left(E_P\left[G(Z)\right] - V_T\right)^2~.
\end{equation}

The algorithm for the determination of the optimal
sampling density within a certain trial family can be therefore
summarized as it follows:
\begin{enumerate}
\item Generate a suitable number $N_p^\prime$ of replications of the
state variables $Z$ according to the {\em original} probability
density $P(Z)$;

\item Choose a trial probability density $\tilde P_\theta(Z)$, and
an initial value of the vector of parameters $\theta $;

\item Set $ x_i\to Z_i$, $f_\theta(x) \to W_\theta(Z)^{1/2} G(Z)$
and $y_i \to 0$ ({\em resp.} $y_i \to V_T$) and call a least squares fitter, say ${\rm
LSQ}\,\left[x,y, f_\theta(X),\theta\right]$~, providing the
optimal $\theta= \theta^\star$ by minimizing the second moment of
the estimator $m_2(\theta)$, Eq.~(\ref{minimizer}) [{\em resp.}
$S_2(\theta)$, Eq.~(\ref{pseudovariance})].

\end{enumerate}

Once the optimal parameters $\theta^\star$ have been determined through the least squares algorithm, one can
perform an ordinary Monte Carlo simulation  by sampling the probability density
$\tilde P_{\theta\star}(Z)$, and calculating expectation values according to Eq.~(\ref{isamp}).

What makes LSIS a practical strategy 
is that just a relatively small number
of replications $N_p^\prime\ll N_p$ is usually required to
determine the optimal parameters $\tilde \theta^\star$. This is due
to the fact that the configurations over which the optimization is
performed are fixed. As a result of this form of {\em correlated
sampling} \cite{umrigar98}, the difference in the $m_2(\theta)$'s
for two sets of values of the parameters being optimized is much
more accurately determined than the values of the $m_2(\theta)$'s
themselves. This rather surprising feature is rooted in the fact
that the minimization of Eq.~(\ref{minimizer}) as a means to
optimize the trial density, $\tilde P_\theta(Z)$, can  be
justified in terms of a genuine maximum likelihood criteria
\cite{Bressanini02}, and it is therefore independent on how
accurately $m_2(\theta)$ approximates the quantity (\ref{second}).
As a result, the overhead associated with the optimization
of the trial density is generally fairly limited, thus making LSIS a practical 
approach for variance reduction.

In a companion paper \cite{LC_lsis06} we have demonstrated the effectiveness
of LSIS by applying it to a variety of test cases. In particular, we 
have shown that LSIS provides variance reductions comparable or superior
to those of the Importance Sampling methods most recently proposed
in the financial literature \cite{GlassImportSampl99,SUFU00,SUFU02,Arouna03}.
As a simple example, for instance, below we briefly review the results obtained
for  standard Call and Put options in a Black-Scholes setting. 
In this case the payout function reads as in Eq.~(\ref{call}) (for the call), 
and the sampling density $P(Z)$ is a univariate 
standard normal density.  

As discussed above,  Importance
Sampling techniques seek a sampling probability density $\tilde
P_\theta(Z)$ as close as possible to the optimal sampling
density, Eq.~(\ref{zerovar}) (see Figure \ref{figcall}). The
simplest choice for $\tilde P_\theta(Z)$, in this setting, is a
Gaussian density of the form (\ref{trialgauss}) (with $d=1$),
so that the only parameter $\theta$ to optimize is the drift
$\tilde\mu$.  We found
that the least squares fitter was able to determine successfully
the optimal $\tilde \mu$ with as little as $N_p^\prime \simeq 50$
Monte Carlo replications.

In Tables \ref{tablecall} and \ref{tableput} we compare the
results obtained with LSIS
with the ones obtained by means of the Robbins Monro (RM) adaptive
Monte Carlo (as quoted in Ref.~\cite{Arouna03}), and the saddle
point approach of GHS \cite{GlassImportSampl99}.
 Here,  as an indicator of the efficiency gains
introduced by the different strategies of Importance Sampling, we have defined the
variance ratio as
\begin{equation}\label{varred}
{\rm VR} = \left(\frac{\sigma({\rm Crude\,\,MC})}{\sigma({\rm IS})}\right)^2
\end{equation}
where the numerator and denominator are the statistical errors 
(for the same number of Monte Carlo paths)
of the Crude and the Importance Sampling estimators, respectively.

We found that the different methods produce a significative and
comparable variance reduction. Intuitively,
the change of drift is more effective for low volatility,
and deep in and out of the money options (see also the discussion in the Introduction). 
In this case, the  LSIS and
GHS optimized trial densities $\tilde P_{\bar \mu}(Z)$ are
very similar as shown Fig.~\ref{figcall}. This could be expected
as, in this case, the optimal Importance Sampling  density
(\ref{zerovar}) can be effectively approximated by a Gaussian with
the same mode and variance, so that the GHS approach produces
accurate results.

However, the LSIS method is not limited to Importance Sampling
strategies based on a pure change of drift, and one can easily
introduce additional optimization parameters  in the trial density.
For instance, in this example it makes sense to introduce the
sampling volatility,  $\tilde \sigma$,
\begin{equation}\label{trialvol}
\tilde P_{\tilde\mu,\tilde\sigma}(Z) = (2 \pi \tilde \sigma^2)^{-1/2}
e^{-(Z-\tilde\mu)^2/2\tilde \sigma^2}~.
\end{equation}
As illustrated in Fig.~\ref{figcall}, by adjusting both $\tilde\mu$ and
$\tilde \sigma$, one obtains a trial density closer to
the optimal one. This corresponds to an additional variance
reduction up to over one order of magnitude, as shown in Tables
\ref{tablecall} and \ref{tableput}.

\section{Stratified Sampling}
\label{ss}

In a diffusive setting, LSIS
can be naturally combined with stratified sampling
\cite{HammersleyHandscomb64}  in order to achieve further variance 
reductions.
In this Section we illustrate how. We begin by reviewing the
basic ideas underlying Stratification following
Refs.\cite{GlassImportSampl99,GlassMCbook}.

Stratification is a technique that allows one to draw samples from
a specified distribution in a more regular pattern thus reducing the variance.
This is achieved by ensuring that the fraction of samples which falls
in different subsets, or {\em strata}, of the domain of the random variable
matches the theoretical probability of each subset. For example, in order
to perform a stratified sampling of a single standard normal variable one can
divide the real axis into $M$ strata, such that the probability of the 
random variable to fall in any of them is $1/M$. This can be done easily
by first dividing the unit interval $(0,1)$ into $M$ segments of length $1/M$,
and sampling uniformly from each of them. Then, each of the sampled uniform
is mapped into a standard Gaussian by means of the inverse cumulative normal 
distribution. The resulting set of $M$ variates will contain exactly one variable for each 
of the $M$ strata of the real axis, and constitute therefore a stratified sample
of the standard normal distribution. This simple algorithm can be therefore summarized as it 
follows:
\begin{enumerate}
\item Draw $M$ random variables, say 
 $u^{1},\ldots,u^{M}$, uniformly distributed in $(0,1)$. 
\item Define a new set of $M$ random variables 
$$ v^{(i)} = \frac{i-1}{M} + \frac{u^{(i)}}{M}~,$$
with $i=1,\ldots, M$, i.e., such that the $i$-th variable is uniformly distributed
in the interval $( i-1/M,i/M )$.
\item Set
$$
X^{(i)} = \Phi^{-1}(v^{(i)})~,
$$
where $\Phi$ is the standard normal cumulative density function. The variables 
$(X^{(1)},\ldots,X^{(M)})$ constitute
the sample of the standard normal distribution, stratified into $M$ strata.
\end{enumerate}

Although this procedure can be generalized to multi-dimensional
normal variates, it becomes unpractical in high-dimension
($d \gtrsim 5$) for the same reason for which estimating the integral
(\ref{expectproblem}) by numerical quadrature becomes exponentially inefficient:
if each dimension is divided into $M$ strata, their total number
scales as $M^d$.  As a result, generating just one point on each stratum
requires a sample size at least this large, thus becoming prohibitive
for the values of $M \gtrsim 10$ that generally make Stratification
effective in reducing the variance. 

 A feasible way of applying Stratification
to the sampling of a multi-variate normal distribution is to stratify only
a specific one-dimensional projection of the random variable $Z \sim N(0,I_d)$.
This is straightforward because, the projection of $Z$ along a
direction in $\mathbb{R}^d$ represented by a unit vector $\xi$, $\xi \cdot Z$, is
a standard normal variable that can be stratified using the one-dimensional
algorithm described above. In addition, it is also
easy to sample the vector $Z$ conditional to a 
specific value of its projection $\xi \cdot Z$, as the conditional distribution
$( Z | \xi \cdot Z = x)$ is itself normal and given by $N (x \xi, I_d - \xi \xi^t)$. 
The resulting algorithm leading to the stratification of $Z$ along the direction
$\xi$ can be therefore summarized as it follows:
\begin{enumerate}
\item Generate a stratified sample of $X^{(1)},\ldots, X^{(M)}$ of the standard
normal distribution as described above. Interpret $X^{(i)}$ as the the $i$-th
value of the one-dimensional projection $\xi \cdot Z$, of $Z \sim N(0,I_d)$.

\item Draw $M$ independent $d$-dimensional Gaussian variates $Y^{(i)}$ from $N(0,I_d)$. 

\item Set 
$$ Z^{(i)} = \xi X^{(i)} + (I_d - \xi \xi^t ) Y^{(i)}~. $$

\end{enumerate}
The resulting set $(Z^{(1)},\ldots,Z^{(M)})$ constitutes a sample from
$N(0,I_d)$ stratified along the direction $\xi$ into $M$ strata.

Loosely speaking, the Stratification of a one-dimensional projection of a multi-dimensional
normal variate has nearly the same effect of replacing the Monte Carlo integration with
a numerical quadrature along the stratified direction $\xi$, while still 
using Monte Carlo for the remaining ones. Clearly, the choice of the
direction $\xi$ is critical for the Stratification to be effective in terms
of variance reduction. This is likely to be the case if the output is
strongly correlated to the value of the projection $\xi \cdot Z$. 

As anticipated, the simplest possible strategy for Importance Sampling in a
Gaussian framework, is to look for an optimal change of drift, i.e. to 
adopt the simple shifted Gaussian of Eq.~(\ref{trialgauss})
as trial probability density. 
In this setting, as suggested by Glasserman and collaborators \cite{GlassImportSampl99},
a natural choice for the direction of stratification is the optimal drift 
vector itself. This can be rigorously justified if the payout 
is a function of a linear combination of the $Z_i$'s. However,
in Refs.\cite{GlassImportSampl99,GlassImportSampl99hjm} and \cite{LC_lsis06} 
it has been shown that this choice works in practice more in general, turning out to be
highly effective in a variety of cases. In this paper, we also follow
this strategy, and demonstrate its effectiveness for a variety 
of examples in the context of the Libor Market Model.

\section{ The Libor Market Model Setting}
\label{lmm}

In the remainder of this paper we will apply the LSIS strategy,
reviewed above, to the Libor Market Model of Brace, Gatarek 
and Musiela \cite{BGM} for the arbitrage-free evolution of the forward Libor rates. In order to 
introduce this framework, we indicate with $T_i$, $i = 1, \ldots, M+1$,
a set of $M+1$ bond maturities, with spacings $h = T_{i+1}-T_i$, assumed constant for simplicity.
The Libor rate as seen at time $t$ for the interval
$[T_i,T_{i+1})$, $L_i(t)$, evolves according to the following
stochastic differential equation
\begin{equation}\label{lmm_sde}
\frac{dL_i(t)}{L_i(t)} = \mu_i(L(t)) dt + \sigma_i(t)^T dW_t, ~~~~ 0\leq t\leq T_i, ~~~ i=1,\ldots,M~, 
\end{equation}
where $W$ is a $N$-dimensional standard Brownian motion, $L(t)$ is the $M$-dimensional 
vector of Libor rates, and $\sigma_i(t)$ the $N$-dimensional vector of volatilities, both at time 
$t$. Here the drift term, as imposed by the arbitrage free conditions, reads
\begin{equation}
\mu_i(L(t)) = \sum_{j=\eta(t)}^i \frac{\sigma_i^T\sigma_j h L_j(t)}{1+ h L_j(t)}~,
\end{equation}
where $\eta(t)$ denotes the index of the bond maturity immediately following time $t$, 
with $T_{\eta(t)-1} \leq t < T_\eta(t)$.  

Equation (\ref{lmm_sde}) can be simulated by applying a Euler discretization to the logarithms of the forward rates, 
and by  dividing each interval $[T_i,T_{i+1})$ into $n_e$ steps of equal width, $h_e = h/n_e$. This gives
\begin{equation}\label{lmm_eul}
\frac{L_i(n+1)}{L_i(n)} = \exp\left[ \left(\mu_i(L(n)) - ||\sigma_i(n)||^2/2\right) h_e 
+ \sigma_i^T(n) Z(n+1) \sqrt{h_e}\right],~
\end{equation}
for $i=\eta(nh_e),\ldots,\ldots,M$, and $L_i(n+1) = L_i(n)$ if $i < \eta(nh)$. Here $Z$ is a
$N$-dimensional vector of independent standard normal variables. 
Under the discretized model (\ref{lmm_eul}), the problem of evaluating the price of a 
contract written on a set of Libor rates is then formulated in the general form (\ref{expectproblem}), 
and LSIS can be straightforwardly applied.

In the following we will present results using a trial probability density 
involving displaced Gaussian multi-variate densities of the form (\ref{trialgauss}).
This choice requires in principle the optimization of a number of parameters -- 
the components of the drift vector $\tilde{\mu}$ -- proportional
to the number of Gaussian univariate $Z_i$ necessary for the propagation of the Libor rates in
the desired time horizon,
namely $d = M\times N \times n_e$. As the number of time steps or the number of factors of 
the simulation  increase, the complexity of the optimization problem increases as well.
Nevertheless, as suggested in Ref.~\cite{GlassImportSampl99hjm} and verified in the companion paper
\cite{LC_lsis06} for a variety of examples, one can significantly reduce the
computation time associated with the optimization stage by
approximating the drift vector with a continuous function
parameterized by a small number of parameters. These are
in turn tuned by the least square algorithm in order to determine
an approximate optimal 
drift vector. We have found that a particularly
effective realization of this approach is to approximate the
drift vector by a piecewise linear function, parameterized by its
values where it changes slope (the so-called {\em knot points}). 
In particular, in the simulation of the LMM we have found that by using
a very limited number of knot points for each random factor (say for 1 to 5)
one is able to achieve very effective variance reductions through LSIS and LSIS plus Stratification.
Hence the simulation of the LMM required the optimization of a very small
number of parameters (form 3 to 15, for $N=3$) thus making  the overhead associated with the
presimulation stage rather limited. More precisely, we found that a few hundred Monte Carlo
configurations and 10-20 iterations of the least squares fitter,
were typically enough to determine the optimal drift vector.  In addition, such vector generally
changes continuously with the simulation parameters.
As a result, an even faster convergence in the iterative procedure can be obtained
by starting the pre-simulation from a drift vector optimized for a case with a similar set of
parameters.

\begin{table*}
\caption{\label{table_caplet}  Variance reductions (\ref{varred}) obtained with LSIS and LSIS plus
Stratification (LSIS+) for Caplets, Eq.~(\ref{caplet}), in a three factor Libor Market Model, for different maturities $T_m$,
and strike prices $K$. $N_k$ is the number of knots per factor (see text).
The number of simulated paths is 200,000.
The uncertainties on the variance reductions are reported in parentheses.
}
\center
\begin{tabular}{ccccc}
$T_m$ (years) &   $K$  &  $N_k$ &  LSIS    & LSIS+  \\
\hline
 1.0 &  0.04  &  1     &  11.4(1) & 1349(1) \\
 1.0 &  0.055 &  1     &  13.3(2) & 2300(2) \\
 1.0 &  0.07  &  1     &  20.2(1) & 4126(4) \\
 2.5 &  0.04  &  1     &  14.0(1) & 1189(1) \\
 2.5 &  0.055  &  1     &  15.5(1) & 897(1)  \\
 2.5 &  0.07  &  1     &  18.1(1) & 1831(1)  \\
 5.0 &  0.040 &  1     &  12.7(1) & 235.2(5) \\
 5.0 &  0.060 &  1     &  12.5(1) & 237.0(5) \\
 5.0 &  0.080 &  1     &  14.5(1) & 193.3(4) \\
 7.0 &  0.04  &  1     &  7.9(3)  & 40.0(1)  \\
 7.0 &  0.055 &  1     &  8.5(4)  & 43.7(1)  \\
 7.0 &  0.07  &  1     &  8.5(4)  & 40(1)   \\
\hline
\end{tabular}
\end{table*}

\section{Numerical Results}
\label{numres}

The numerical results we present in this Section are based on the evolution
of (\ref{lmm_eul}) in a three-factor ($N=3$) model with $h = 1/4$ (a quarter 
of a year), and  $n_e = 3$.  Following Ref.~\cite{GlassZhao99}, to keep things simple 
we take the volatilities to be functions of time to maturity 
\begin{equation}
\sigma_i(t) = \sigma_{i-\eta(t)+1}(0)~,
\end{equation}
with 
\begin{equation}
\sigma_i^j(0) = \sigma_0 (1 + \alpha j)(1 + \beta i)~,  
\end{equation}
$j=1,\ldots,3$, $\alpha=0.1$ and $\beta=0.01$, and $\sigma_0 = 0.2$.  As initial Libor curve we take
instead 
\begin{equation}
L_i(0) = l_0 (1 + \beta i)~,
\end{equation}
with $l_0 = 5\%$.

As a first example we consider a Caplet for the interval $[T_m,T_{m+1})$ struck at $K$, 
\begin{equation}\label{caplet}
C_h(T_m) = \left(\prod_{i=0}^m \frac{1}{1 + h L_i(T_i)}\right) h (L_m(T_m)-K)^+~.
\end{equation}
Table \ref{table_caplet} displays the estimated variance ratios obtained with LSIS, and the combination
of LSIS and Stratification (LSIS+) introduced in Section \ref{ss} 
for a variety of maturities, and strike prices that range from in the money to out of the money. 
Here the results are all obtained using (\ref{trialgauss}) as trial probability density, 
and by parameterizing the change of drift of each factor with 
a single parameter or knot point, corresponding to a rigid shift. We have verified that increasing 
the number of knots does not provide further sizable benefits in this case. As shown in Table \ref{table_caplet},
LSIS provides remarkable variance reductions, corresponding to a 
saving of roughly one order of magnitude in computational
time, consistently across maturities. For fixed
maturity, as expected, LSIS is more effective for out of the money strikes
since in these cases the fraction of paths expiring worthless is more significant. 
These paths clearly provide 
little information, and tend to increase the variance of the sample. Changing the drift 
increases the fraction of paths which end up in the money thus making the sample more homogeneous.
Conversely, as the maturity increases, the variance reduction provided by LSIS decreases
as the outturn distributions of the Libor rates become more delocalized, and the change of drift
strategy becomes less effective. 

The combination of LSIS and Stratification provides for Caplets a tremendous variance reduction
of up to two orders of magnitude (see Table \ref{table_caplet}). 
However, the effectiveness of LSIS+ decreases sharply with maturity. Nevertheless,
for the examples considered, it still gives around a factor of 40 in
variance reduction for a 7 year maturity, thus resulting in extensive savings in computational time
also for fairly long expiries.

\begin{table*}
\caption{\label{table_cap}  Variance reductions obtained with LSIS and LSIS plus
Stratification (LSIS+) for Caps Eq.~(\ref{cap}) in a three factor Libor Market Model, for $T_n=0.25$ (years),
different final maturities $T_M$,
and strike prices $K$. $N_k$ is the number of knots per factor (see text).
The number of simulated paths is 200,000.
The uncertainties on the variance reductions are reported in parentheses.
}
\center
\begin{tabular}{ccccc}
$T_M$ (years) &   $K$  &  $N_k$ &  LSIS    & LSIS+  \\
\hline
 1.0 &  0.04  &  3     & 10.6(5) & 37.2(8) \\
 1.0 &  0.055 &  3     & 9.7(3)  & 19.8(5) \\
 1.0 &  0.07  &  3     & 13.6(5) & 21.6(6) \\
 2.5 &  0.04  &  3     & 16.2(5) & 40.3(7) \\
 2.5 &  0.055 &  3     & 12.0(4) & 33.8(7) \\
 2.5 &  0.07  &  3     & 15.7(5) & 47.3(8) \\
 5.0 &  0.04  &  3     & 14.9(5) & 43.7(9) \\
 5.0 &  0.055 &  3     & 14.5(6) & 46.7(9) \\
 5.0 &  0.07  &  3     & 15.6(6) & 55(1)   \\
 7.0 &  0.04  &  3     & 13.0(6) & 42.6(8) \\
 7.0 &  0.055 &  3     & 12.2(5) & 45.1(9) \\
 7.0 &  0.07  &  3     & 12.6(4)   & 55(1)   \\
\hline
\end{tabular}
\end{table*}

Although important instruments for calibration, Caplets
constitute an easy test ground for LSIS and LSIS+ as they are mostly sensitive to the single Libor
rate determining the final payment. A more articulated example on which to assess the efficacy 
of LSIS are interest rate Caps. We consider 
contracts with first payment $T_n$ and last payment $T_M$, and tenor $h$
\begin{equation}\label{cap}
Cap_h(T_n, T_M) = \sum_{l=n}^M C_h(T_l)~.
\end{equation}
The results obtained for a variety of maturities and strike prices are shown in Table \ref{table_cap}. 
In this case we have verified that $N_k = 3$ knot points
provided the bulk of the variance reduction for the trial density function (\ref{trialgauss}).
The efficiency gains produced by LSIS, although slightly smaller than in the case of a single
Caplet, are consistently around $10-15$ for all the maturities considered. As expected,  LSIS+ is
not able to provide the massive variance reductions observed for Caps. Nonetheless,
for the cases considered, it provides a further reduction of the variance with respect to LSIS 
of a sizable factor ranging from 2 to 4.

\begin{table*}
\caption{\label{table_swaption}  Variance reduction obtained with LSIS and LSIS plus
Stratification (LSIS+) for Swaptions Eq.~(\ref{swaption}) in a three factor Libor Market Model.
$T_n$ is the option expiry and $T_{M+1}$ is the final payment date of the underlying swap. $K$ is the
strike price. $N_k$ is the number of knots per factor (see text).
The number of simulated paths is 200,000.
The uncertainties on the variance reductions are reported in parentheses.
}
\center
\begin{tabular}{cccccc}
$T_n$ (years) & $T_{M+1}$ & $K$  &  $N_k$ &  LSIS    & LSIS+  \\
\hline
 0.5        & 1.5   & 0.04  &  3     &   6.8(3)   & 35.2(8) \\
 0.5        & 1.5   & 0.055 &  3     &   10.5(4)  & 143(2)  \\
 0.5        & 1.5   & 0.07  &  3     &   21.2(6)  & 209(2)  \\
 0.5        & 2.5   & 0.04  &  3     &   7.0(3)   & 41.9(9) \\
 0.5        & 2.5   & 0.055 &  3     &   9.8(3)   & 149(2)  \\
 0.5        & 2.5   & 0.07  &  3     &   18.6(5)  & 427(2)  \\
 0.5        & 5.5   & 0.04  &  3     &   6.8(3)   & 50(1)   \\
 0.5        & 5.5   & 0.055 &  3     &   8.5(3)   & 106(1)  \\
 0.5        & 5.5   & 0.07  &  3     &   12.0(4)  & 148(1)  \\
 1.0        & 6.0   & 0.04  &  3     &   8.0(4)   & 144(2)  \\
 1.0        & 6.0   & 0.055 &  3     &   8.6(3)   & 165(2)  \\
 1.0        & 6.0   & 0.07  &  3     &   12.7(4)  & 654(3)  \\
 2.0        & 7.0   & 0.04  &  3     &   9.2(3)   & 70(1)   \\
 2.0        & 7.0   & 0.055 &  3     &   9.7(3)   & 139(1)  \\
 2.0        & 7.0   & 0.09  &  3     &   13.9(4)  & 140(1)  \\
 5.0        & 10.0  & 0.04  &  5     &   7.3(4)   & 76(1)   \\
 5.0        & 10.0  & 0.055 &  5     &   7.4(3)   & 72(2)   \\
 5.0        & 10.0  & 0.09  &  5     &   7.5(4)   & 197(2)  \\
\hline
\end{tabular}
\end{table*}

LSIS and LSIS+ result in remarkable computational savings also for Swaptions. Here we 
have considered contracts with expiry $T_n$ to enter in a swap with
payments dates $T_{n+1},\ldots,T_{M+1}$, with the holder of the option paying
a fixed rate $K$
\begin{equation}\label{swaption}
V(T_n) = \sum_{i = n+1}^{M+1} B(T_n,T_i) h (S_n(T_n)-K)^+~,
\end{equation}
where $B(T_n,T_i)$ is the price at time $T_n$ of a bond maturing at time $T_i$
\begin{equation}
B(T_n,T_i) = \prod_{l=n}^{i-1} \frac{1}{1 + h L_l(T_l)},
\end{equation}
and the swap rate reads
\begin{equation}
S_n(T_n) = \frac{1 - B(T_n, T_{M+1})}{h\sum_{ l = n + 1}^{M+1} B(T_n,T_l)}~.
\end{equation}
The results are shown in \ref{table_swaption} and indicate that LSIS provides
variance reductions in the range $7\div 20$ and LSIS+ further increases the
computational efficiency by up to one order of magnitude.

As a final example -- illustrating for a simple case the flexibility of LSIS -- 
we consider the  combination of a long Caplet
and Flooret in a Straddle contract
\begin{equation}\label{straddle}
St_h(T_m) = \left(\prod_{i=0}^m \frac{1}{1 + h L_i(T_i)}\right) h |L_m(T_m)-K|~.
\end{equation}

In this case, the  optimal sampling density (see Sec.\ref{importance}), proportional to
the product of the payout and the Gaussian sampling density (\ref{gaussmulti}),
has two well separated maxima because of the modulus in Eq.~(\ref{straddle}). As a result, a single mode trial probability density
(\ref{trialgauss}) provides limited variance reductions, especially for
strikes at the money, where the relative importance of the two maxima is
similar (see Tab.~\ref{table_straddle}). However,  the LSIS is not limited to 
a Gaussian trial density and one can use this flexibility to utilize
a more accurate guess of the optimal sampling density. In particular,  
a  better {\em ansatz} for
the optimal density is represented by a bi-modal trial
density of the form
\begin{equation}\label{bimodal}
\tilde P(Z) = (2\pi)^{-d/2}\Big[w_a \,e^{-(Z-\mu_a)^2/2}+w_b\,
e^{-(Z-\mu_b)^2/2}\Big]~,
\end{equation}
where $w_a+w_b=1$ that can be
optimized over $\mu_a$, $\mu_b$, and $w_a$. The simulation of a
density of this form is straightforward as it simply implies
choosing one of the two Gaussian
components in (\ref{bimodal})  on each Monte Carlo step, and sample a configuration $Z_i$
according to it. This can be done by extracting an auxiliary
uniform random number $\xi \in [0,1]$, and sampling $Z_i$
according to the first Gaussian component if $\xi<w_a$, and
according to the second otherwise. 
As shown in Table \ref{table_straddle}, using this trial density,
LSIS improves significantly the computational efficiency also
for Straddle contracts.

\begin{table*}
\caption{\label{table_straddle}  Variance reduction obtained with LSIS
for a Straddle Eq.~(\ref{straddle}) in a three factor Libor Market Model,
for different maturities $T_m$, and strike prices $K$. $N_k$ is the number
of knots per factor (see text). Results are shown using Eq.~(\ref{trialgauss}) [LSIS]
and Eq.~(\ref{bimodal}) [LSIS (MM)] as trial densities.
The number of simulated paths is 200,000.
The uncertainties on the variance reductions are reported in parentheses.
}
\center
\begin{tabular}{ccccc}
$T_m$ (years) & $K$  &  $N_k$ &  LSIS    & LSIS (MM) \\
\hline
 1.0        & 0.04  &  1    &   2.8(1) & 5.8(1) \\
 1.0        & 0.05  &  1    &   1.3(1) & 5.3(1) \\
 1.0        & 0.06  &  1    &   1.0(1) & 3.9(1) \\
 1.0        & 0.07  &  1    &   1.1(1) & 3.4(1) \\
 5.0        & 0.04  &  1    &   2.8(1) & 8.7(1) \\
 5.0        & 0.05  &  1    &   1.9(1) & 6.5(1) \\
 5.0        & 0.06  &  1    &   1.5(1) & 4.9(1) \\
 5.0        & 0.07  &  1    &   1.2(1) & 4.0(1) \\
\hline
\end{tabular}
\end{table*}

\section{Conclusions}
\label{conclusions}

In this paper we have described 
the application of the recently introduced Least Squares 
Importance Sampling  (LSIS) \cite{LC_lsis06}
to the simulation of Libor Market Models. 
Such variance reduction technique allows one to automatically
optimize the sampling density within a chosen trial
class by means of a presimulation algorithm of straightforward
implementation.  

What makes the approach practical in a financial context is that the
overhead associated with the least squares optimization of the trial density is 
generally rather limited especially after reducing
the dimensionality of the problem by means of a careful parametrization. 

With several numerical examples we have shown that LSIS can be extremely effective in
reducing the variance per sample of the simulation, thus resulting in remarkable
speed-ups. Moreover,  when used with Gaussian trial probability densities, LSIS can be
naturally combined with Stratification thus providing further efficiency
gains that can result in computational savings of orders of magnitude.

The efficacy of any Importance Sampling strategy is much dependent on 
how effectively the trial density function is able to reweigh the different regions of the
sampled space in order to reduce the statistical fluctuations of the accumulated 
observables. These regions depends on both the model simulated, and the structure of the 
payout being priced. In this respect LSIS, when compared with previously methods, 
offers additional potential leeway as it is not limited to Gaussian trial densities. 
This becomes important when the structure of the optimal density  is particularly complex
e.g., with multi-modal features, or complicated  correlation structures.
In this paper we have illustrated this point with a simple multi-modal example. Further work is
currently in progress in order to introduce more flexible probability distributions 
as trial densities.

{\bf Acknowledgments:} It is a pleasure to acknowledge 
Gabriele Cipriani,
David Shorthouse, and Mark Stedman for stimulating
discussions, and Paul Glasserman for kind and useful correspondence. 
The opinion and views 
expressed in this paper are uniquely those of the author, and do 
not necessarily represent those of Credit Suisse Group.

\bibliographystyle{plain}
\bibliography{biblio}% Produces the bibliography via BibTeX.

\end{document}